Fabrication of single nickel-nitrogen defects in diamond by chemical vapor deposition

J. R. Rabeau,[a)] Y. L. Chin,[b)] and S. Prawer

*School of Physics, University of Melbourne, Victoria, Australia, 3010*

F. Jelezko, T. Gaebel, and J. Wrachtrup

*Physikalisches Institut, Universitat Stuttgart, Pfaffenwaldring 57, 70569 Stuttgart, Germany*

a) corresponding author email: jrabeau@unimelb.edu.au

b) current address: *Electrical & Computer Engineering, University of Canterbury, Private Bag 4800, Christchurch, New Zealand*

Abstract

Fabrication of single nickel-nitrogen (NE8) defect centers in diamond by chemical vapor deposition is demonstrated. Under continuous-wave 745 nm laser excitation single defects were induced to emit single photon pulses at 797 nm with a linewidth of 1.5 nm at room temperature. Photon antibunching of single centers was demonstrated using a Hanbury-Brown and Twiss interferometer. Confocal images revealed approximately $10^6$ optically active sites/$cm^2$ in the synthesized films. The fabrication of an NE8 based single photon source in synthetic diamond is important for fiber based quantum cryptography. It can also be used as an ideal point-like source for near-field optical microscopy.

Optically active defects in solids offer many potential uses in quantum computing and cryptography. Diamond, for example, has optically active color centers which can be used for on-demand single photon generation. It is the only material presently known which contains room temperature photostable defects capable of producing single photon pulses. The nitrogen-vacancy (N-V) color center in diamond has already been demonstrated to be an efficient source for single photons[1] and has been used to implement quantum key distribution in free space.[2] It is also being explored as a spin qubit in quantum computing.[3]

Recently, a nickel-related defect in diamond was identified to be an efficient source of single photons in the infrared.[4] This defect was assigned to the nickel-nitrogen complex (NE8) in diamond.[5] Electron paramagnetic resonance studies have shown this center to comprise of a substitutional nickel atom, with four adjacent nitrogen atoms.[5,6,7] The NE8 color center emits at ~800 nm, has a room temperature linewidth of 1.2 nm and a lifetime of 11 ns. In the context of a single photon source for fiber-optic communications, the NE8 center displays properties which are in many respects superior to the N-V center: the attenuation of standard silica glass optical fiber at the N-V wavelength (637 nm to 740 nm) is ~7dB/km, whereas for the NE8 wavelength (800 nm) it is ~2.8 dB/km.[8] Furthermore, the NE8 linewidth is only ~1.5 nm at room temperature, compared to the very broad spectral emission of ~100 nm from the N-V center. Emission from the NE8 center therefore experiences approximately 3 orders of magnitude less dispersion broadening in standard optical fiber than N-V emission.[9] This is an especially important parameter when considering communications over long distance. Despite these highly desirable characteristics, no controlled fabrication of this optical centre in diamond has been reported.

The formation energy of nitrogen related complexes in diamond is such that it is very favorable for nitrogen to be incorporated into the diamond lattice. Hence nearly all natural and synthetic diamond contains some background level of N, and as a result most diamonds contain at least some background concentration of N-V centers. The presence of such a background poses a challenge if single N-V centers are to be fabricated in a controlled fashion. In the case of CVD diamond, unless great care is taken to remove traces of residual N (both from the gas feedstock and any residual leaks), a background of N-V centers is often present. By contrast, there is no source of Ni in a typical CVD deposition chamber, and hence the background concentration of Ni and related optical centers in CVD diamond is essentially zero. In particular the Ni related NE8 center, which has been reported in some natural type IIa diamond and high-pressure high-temperature diamond where nickel has been used as a catalyst, has not yet been reported in CVD diamond.

In this paper a method to fabricate single NE8 defects in diamond thin films using chemical vapor deposition (CVD) is reported. The ability to fabricate this center in a controlled way, particularly the fabrication of single, isolated, optically active defects, opens numerous possibilities for photophysical characterization and implementation of in-fiber single photon sources and integrated quantum photonics.

Diamond films were grown on 1 × 1 $cm^2$ fused silica substrates using a 1.2 kW microwave plasma chemical vapor deposition reactor (ASTeX). The chamber pressure was maintained at 30 Torr with a 0.7% $CH_4$ in $H_2$ gas mixture. The substrate temperature was a constant 700ºC during the growth period. For the growths reported here, nitrogen was not deliberately added to the gas feedstock as it is known to be present at a background level of ~0.1%, which corresponds to an N/C ratio of 0.15.

Prior to growth, the fused silica substrates were seeded by exposure to a slurry consisting of nickel (Riedel-de Haen) and diamond powder (<10 nm, De Beers) in an ultrasonic bath. This served to seed the substrate with nano-particles of Ni and diamond. Previous work[10] employing alumina and diamond slurries showed that the nucleation density for diamond could be increased by increasing the alumina particle size (up to 45 μm). Sonochemical synthesis has also been shown to produce nanometer sized metal colloid particles.[11,12] In our work, it was anticipated that combining nickel with diamond would have the effect of nucleating the silica substrate with nano-particles of diamond and nickel, thus facilitating both high nucleation density on the silica and incorporation of Ni in the growing diamond. After the nucleation procedure, the samples were rinsed in acetone and de-ionized water, loaded into the CVD reactor and diamond was grown for 4 hours. The growth rate was less than 1 μm/hour. For the purpose of this study, two samples were ultimately characterized, each seeded with a different Ni:diamond suspension concentration.

Detection of single NE8 defects was performed using a room-temperature confocal microscope with a 100 X oil immersion objective lens (numerical aperture 1.4). The excitation laser was an Ar-ion pumped Titanium:Sapphire laser operating at 745 nm. A holographic notch filter at 745 nm was used to block the pump beam and an 800 nm ± 5 nm bandpass filter was used to isolate the fluorescence from the NE8 defects in the confocal and interferometer arms for the antibunching measurements. The spectral scans collected the total fluorescence output (i.e. the 800 nm bandpass filter was removed for the spectral scans).

The confocal image in FIG. 1a shows a 16 by 16 $\mu m^2$ scan of the diamond film with a single bright spot and FIG. 1b shows a scanning electron micrograph (SEM) of a different area of the same diamond film at a similar magnification. Further characterization of this film revealed an estimated concentration of $10^6$ optically active NE8 defects/$cm^2$ in this region of the spectrum. Several bright spots were selected separately and fluorescence spectra were collected using a 0.3 m imaging spectrometer (Acton Research). FIG. 2 shows the spectrum at room temperature from the optically active centre highlighted in FIG. 1. The peak at 797 nm is the NE8 zero phonon line (ZPL) and the weak phonon sideband extends to ~840 nm. The Debye-Waller factor of this center, which is a measure of the ZPL intensity relative to the entire band intensity, is 0.7 (compared to 0.04 for the N-V center).

Having identified the NE8 center, the photoluminescence was then directed into a Hanbury-Brown and Twiss interferometer to evaluate the photon statistics. The second order autocorrelation function, $g^{(2)}(\tau)=\langle I(t)I(t+\tau)\rangle/\langle I(t)^2\rangle$ represents the normalized probability to detect the photon pairs spaced by time $\tau$. For zero delay time the autocorrelation function is the probability of emission of two simultaneous photons. For a coherent (Poissonian) light source the autocorrelation function is a constant function with value of unity. FIG. 3 shows the background corrected autocorrelation function with a clear drop at $\tau=0$. The measured sub-Poissonian photon statistics indicated that the fluorescence was originating from a single defect. This result originates from the fact that the emission of a photon by a single quantum system projects the system into the ground state and a second photon can not be emitted simultaneously. The data was normalized and background corrected according to the procedure outline in ref [13]. The background corrected contrast of the antibunching dip was measured to be ~0.6. Note that for N emitters the contrast scales as 1/N. Ideally, the contrast of unity would be expected for single defect.

Deviation of the measured value of contrast from unity is related to following experimental details. First, the width of the antibunching is comparable with the time jitter of the photodetectors. Second, it is difficult to accurately estimate the contribution to the autocorrelation measurements of the relatively strong background fluorescence signal. The non-NE8 background photoluminescence in CVD diamond is most likely due to impurities incorporated in the diamond grain-boundaries. These impurities usually consist of $sp^2$ carbon and/or nitrogen-related defects. Steps can be taken to minimize background luminescence by optimizing the growth parameters and employing nitrogen-getters in the gas flow lines.

The single defect was further characterized for photon rate information by measuring the fluorescence as a function of excitation power. Given a single defect with a finite lifetime, the fluorescence will saturate while the background will continue to rise linearly. FIG. 4 shows the raw data saturation curve, the linear background contribution and the background corrected saturation curve. Because the background fluorescence is relatively strong, its total contribution is prevalent in the saturation data where the curve continues to increase even after saturation. The background corrected data in FIG. 4 was fit using $R=R_\infty(I/I_s)/(1+I/I_s)$ where $I$ is the laser power, $R$ is the fluorescence, $I_s$ is the saturation laser power and $R_\infty$ is the fluorescence corresponding to infinite laser power. Taking into consideration the previously estimated total detection efficiency of the system of 0.5% plus an added factor of 2 loss in transmission through the 800 nm bandpass filter, the maximum photon emission rate was found to be 15.2 x $10^6$ counts $s^{-1}$. This is in very good agreement with the results previously reported by Gaebel *et al.* for the NE8 defect in single crystal diamond.

For the purpose of evaluating the effect of Ni in the slurry used to seed the diamond growth, a second film was grown with a smaller proportion of Ni:diamond in the ultrasonic slurry. This film was characterized in the same way with the confocal microscope and no brightly fluorescing centers were observed in a 40 by 40 $\mu m^2$ scan. This indicated that the density of NE8 incorporation is a function of Ni-doping. Further investigation of the effect of Ni:diamond seeding must be carried out before a thorough understanding and control over single defect concentrations can be achieved. Another parameter not studied in detail so far is the effect of the N content in the feedstock which is likely to be important in the formation of NE8 centers.

In summary, we have demonstrated the effective incorporation of isolated, single optically active NE8 defects into CVD diamond at ~$10^6$ defects $cm^{-2}$. The defects were shown to produce antibunched single photon pulses under 745 nm laser excitation. The use of substrate ultrasonification using a diamond and nickel slurry served the dual purpose of nucleating the substrate for diamond growth and generating nano-scale particles of nickel which facilitated the formation of the NE8 defect. This advancement in materials processing will open many possibilities in the area of quantum information processing, particularly in developing single photon sources for fiber based quantum communications.

Acknowledgements

We thank David Jamieson, Brant Gibson, Alon Hoffman and Andrew Greentree for very useful discussion and Sergey Rubanov for SEM characterization. This work was funded by DARPA QuIST and the Australian Research Council and DFG under the framework of 'Quanten-Informationsverarbeitung'.

Figure captions

FIG. 1a. A 16 by 16 μm$^2$ confocal fluorescence scan of diamond at an excitation wavelength of 745 nm and through an 800 ± 5 nm bandpass filter. The scale bar at the side of the image gives an indication of the relative fluorescence intensity from this region of the diamond sample.

FIG. 1b. A scanning electron micrograph of the same diamond film. This region is not necessarily the same region shown in the confocal image.

FIG. 2. Room temperature photoluminescence spectrum of the defect indicated in FIG 1.

FIG 3. Normalized and background corrected autocorrelation function from a single NE8 defect. The fit represents the following function: $y = 1+c_1 e^{(-x/t1)}-c_2 e^{(-x/t2)}$, where $c_1$ and $t_1$ are characterized by the bunching term and $c_2$ and $t_2$ by the antibunching term. The bunching term is related to the population of a metastable state (for details see reference 4).

FIG. 4. Saturation of the fluorescence from an NE8 defect in CVD diamond (circles). The intensity continues to increase linearly even after saturation due to the relatively strong background luminescence. The squares correspond to the background corrected fluorescence which should account solely for the single defect emission. The triangles correspond to the linear background fluorescence.

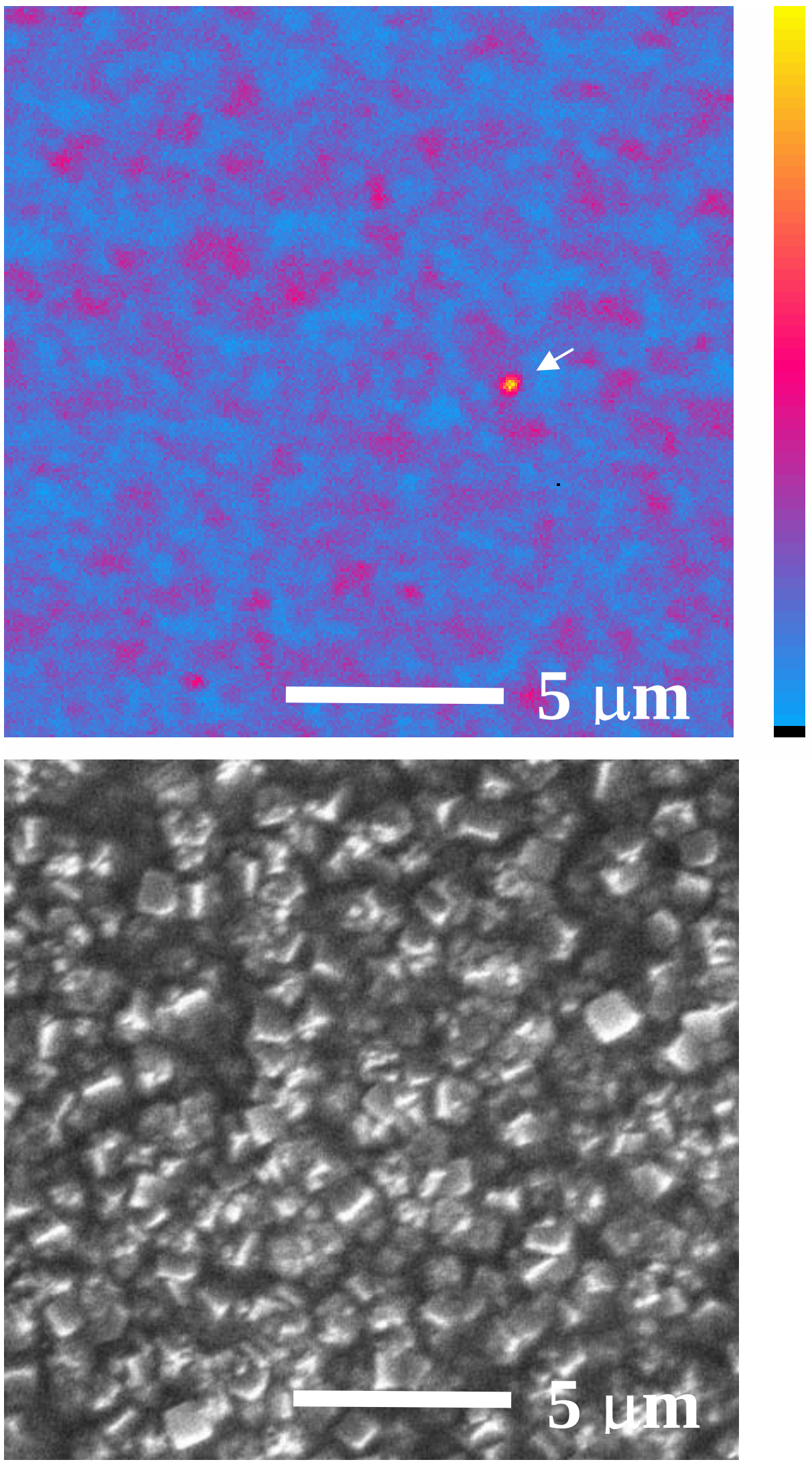

FIG 1.  J. R. Rabeau

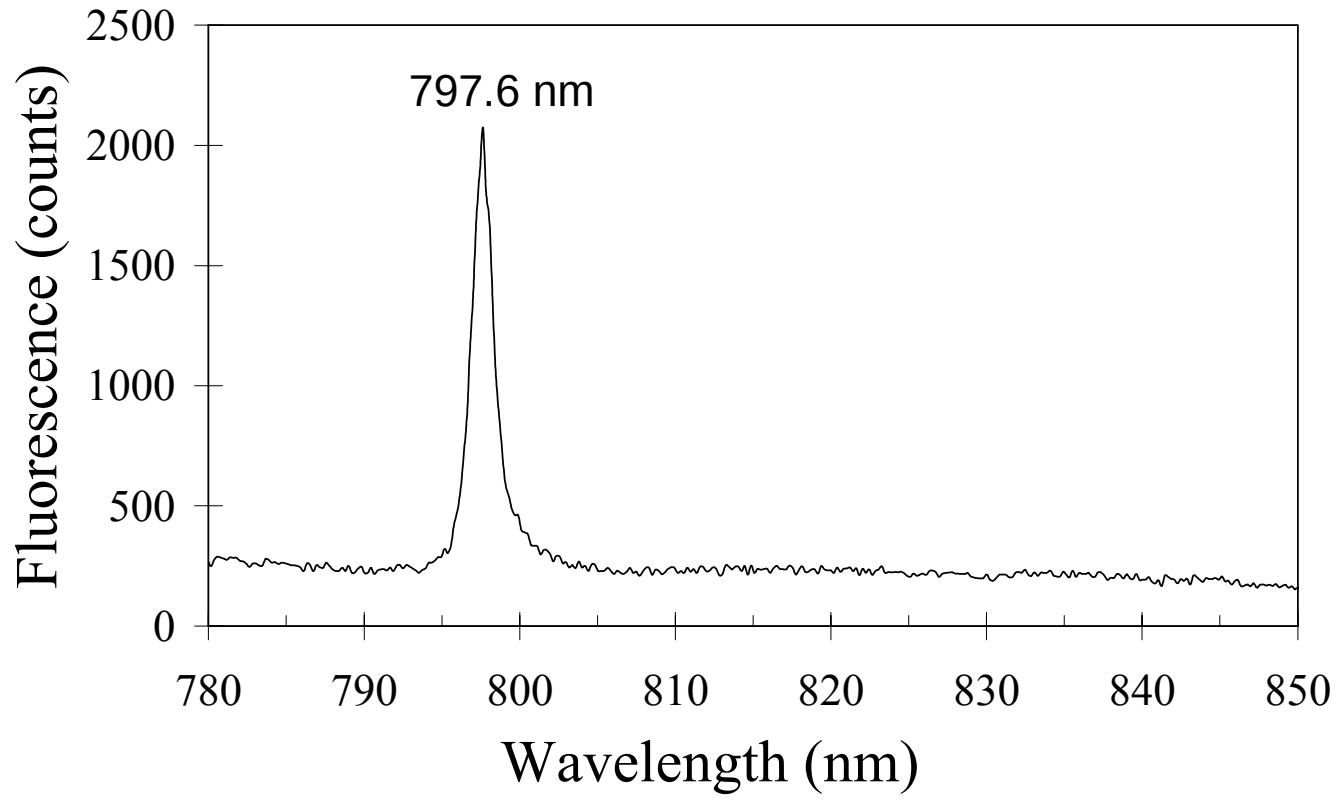

FIG. 2. J. R. Rabeau

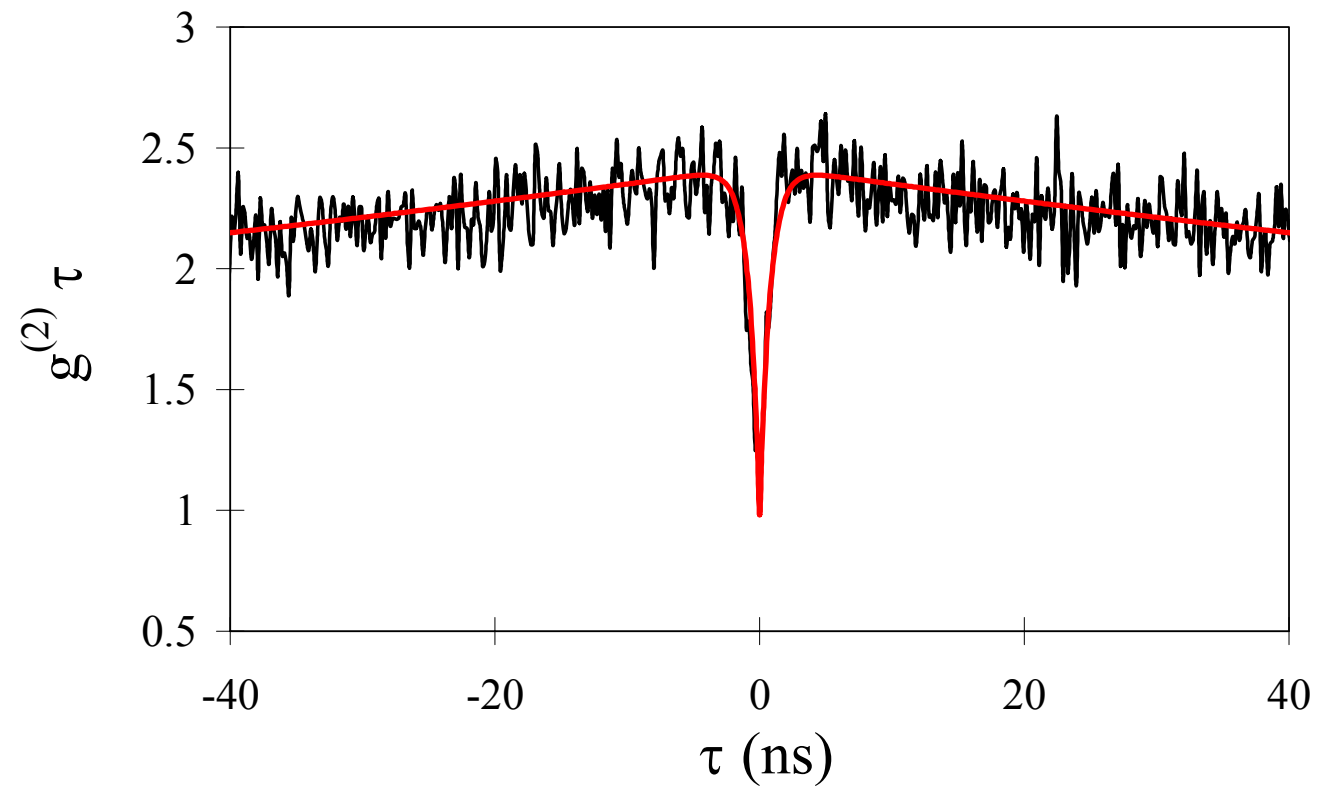

FIG 3. J. R. Rabeau

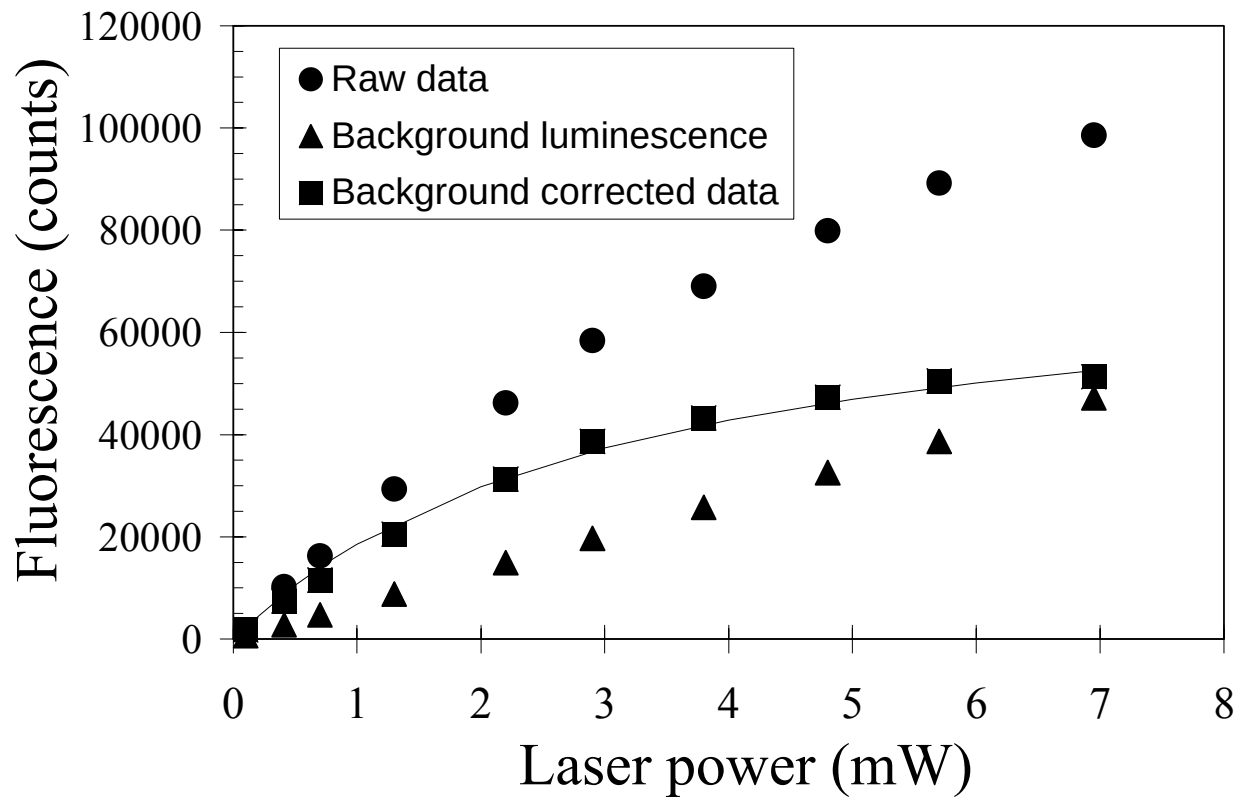

FIG. 4. J. R. Rabeau